\documentclass[]{spie}  

\usepackage{amsmath,amsfonts,amssymb}
\usepackage{graphicx}
\usepackage{xcolor}

\usepackage[colorlinks=true, allcolors=blue]{hyperref}

\title{On-sky, real-time optical gain calibration on MagAO-X using incoherent speckles}

\author[a,b]{Eden A. McEwen}
\author[b]{Jared R. Males}
\author[a,b,c,d]{Olivier Guyon}
\author[e]{Sebastiaan Y. Haffert}
\author[f]{Joseph D. Long}
\author[b]{Laird M. Close}
\author[b]{Kyle Van Gorkom}
\author[a]{Jennifer Lumbres}
\author[a]{Alexander D. Hedglen}
\author[a]{Lauren Schatz}
\author[a]{Maggie Y. Kautz}
\author[b]{Logan A. Pearce}
\author[a,b]{Jay Kueny}
\author[a]{Avalon L. McLeod}
\author[b]{Warren B. Foster}
\author[b]{Jialin Li}
\author[b]{Roz Roberts}
\author[g]{Alycia J. Weinburger}

\affil[a]{James C. Wyant College of Optical Sciences, University of Arizona}
\affil[b]{Steward Observatory, University of Arizona}
\affil[c]{Subaru Telescope, National Observatory of Japan, Hilo, HI}
\affil[d]{Astrobiology Center, National Institutes of Natural Sciences, Japan}
\affil[e]{Leiden Observatory, Leiden University}
\affil[f]{Center for Computational Astrophysics, Flatiron Institute, NY}
\affil[g]{{Earth and Planets Laboratory, Carnegie Institution for Science}}

\authorinfo{Further author information: (Send correspondence to E.A.M.)\\E.A.M.: E-mail: edenmcewen@arizona.edu}

\begin{document} 
\maketitle

\begin{abstract}
The next generation of extreme adaptive optics (AO) must be calibrated exceptionally well to achieve the desired contrast for ground-based direct imaging exoplanet targets. Current wavefront sensing and control system responses deviate from lab calibration throughout the night due to non linearities in the wavefront sensor (WFS) and signal loss. One cause of these changes is the optical gain (OG) effect, which shows that the difference between actual and reconstructed wavefronts is sensitive to residual wavefront errors from partially corrected turbulence. This work details on-sky measurement of optical gain on MagAO-X, an extreme AO system on the Magellan Clay 6.5m. We ultimately plan on using a method of high-temporal frequency probes on our deformable mirror to track optical gain on the Pyramid WFS. The high-temporal frequency probes, used to create PSF copies at 10-22 $\lambda /D$, are already routinely used by our system for coronagraph centering and post-observation calibration. This method is supported by the OG measurements from the modal response, measured simultaneously by sequenced pokes of each mode. When tracked with DIMM measurements, optical gain calibrations show a clear dependence on Strehl Ratio, and this relationship is discussed. This more accurate method of calibration is a crucial next step in enabling higher fidelity correction and post processing techniques for direct imaging ground based systems.  
\end{abstract}

\keywords{Adaptive Optics, Pyramid WFS, Optical Gain, Calibraiton}

\section{INTRODUCTION}
\label{sec:intro}  

Ground based observatories interested in high contrast imaging compensate for atmospheric turbulence with sophisticated adaptive optics (AO) systems. Recent and future high contrast systems have incorporated the Pyramid Wavefront Sensor (PyWFS) for its increased sensitivity. A pyramid WFS \cite{Ragazzoni1996} splits the PSF on the tip of a pyramid into 4 pupil planes, as opposed to the widely used Shack Hartman WFS, which takes a pupil plane and splits into subapertures to sense local slope. The PyWFS provides increased sensitivity in lower order modes but suffers from a degradation of response to, and thus reconstruction of, wavefronts when operated on sky. This loss in the WFS sensitivity is known as Optical Gain (OG), and is an unknown in control systems that limit the future of AO control, from predictive control, wavefront reconstruction, to post-processing techniques. Recent work \cite{VDeo2019, VChambouleyron2020} for next generation ELTs show the importance of compensating for OG through various algorithms in the next generation of ground based instruments. Real time measurement and application of optical gain will allow other systems to reach their full potential for high contrast correction.

\begin{figure}[ht]
    \centering
    \includegraphics[width=0.98\textwidth]{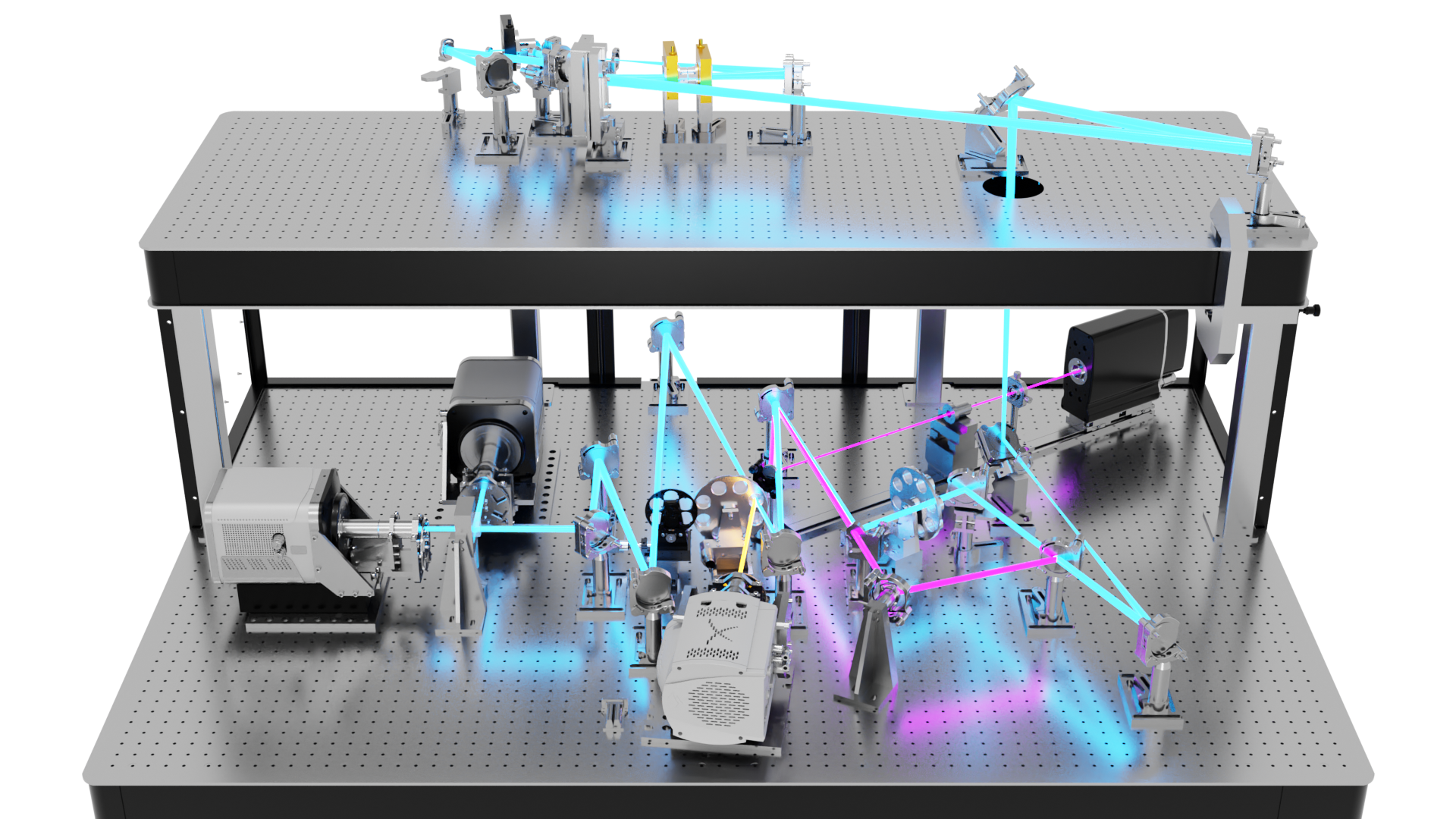}
    \caption{The MagAO-X opto-mechanical design, with the optical path highlighted in blue, and the pyramid WFS path highlighted in purple.}
    \label{fig.magaox_render}
\end{figure}

MagAO-X\cite{males2022magaox}, shown in Figure \ref{fig.magaox_render} is a visible-NearIR extreme AO system built with a visible light pyramid WFS particularly susceptible to the OG effect. The deformable mirror (DM) correction architecture is a woofer-tweeter\cite{Brennan_2006} system, with a high-stroke 97 actuator woofer followed by a 2040 actuator tweeter DM. The visible PyWFS operates modulated, up to 3.6kHZ, though typically at 2kHz modulated at 3 $\lambda /D$. The control architecture is based off of CACAO\cite{Guyon2018} and routinely corrects up to 15 modes. MagAO-X is optimized for high contrast, faint target systems, and is currently being used as a pathfinder for the GMT system. 

In the following, we discuss the work done with the MagAO-X instrument to characterize and calibrate optical gain on sky at the MagAO-X instrument. Though directly recovering the modal response matrix through self response matrices, discussed in section \ref{sect:selfrm}, we are able to get a modal measure of Optical Gain. In Section \ref{sect:OG} we discuss the implications of these measurements and in Section \ref{sect:speckles} we discuss how these OG findings are being used for real time control.

\section{Self Response Matricies}
\label{sect:selfrm}


For on-sky measurement of optical gain per mode, we elected to use the self response matrix (selfRM) a built-in CACAO method for acquiring the modal response matrix. The selfRM probes each mode in the command matrix at the specified amplitude, recording the modal response over time, returning a matrix of size [N modes x N modes x time]. The diagonal gives the modal response, a value close to 1 for lab responses, while the off diagonal terms are cross talk between modes.

\begin{figure}[ht]
    \centering
    \includegraphics[width=0.85 \textwidth]{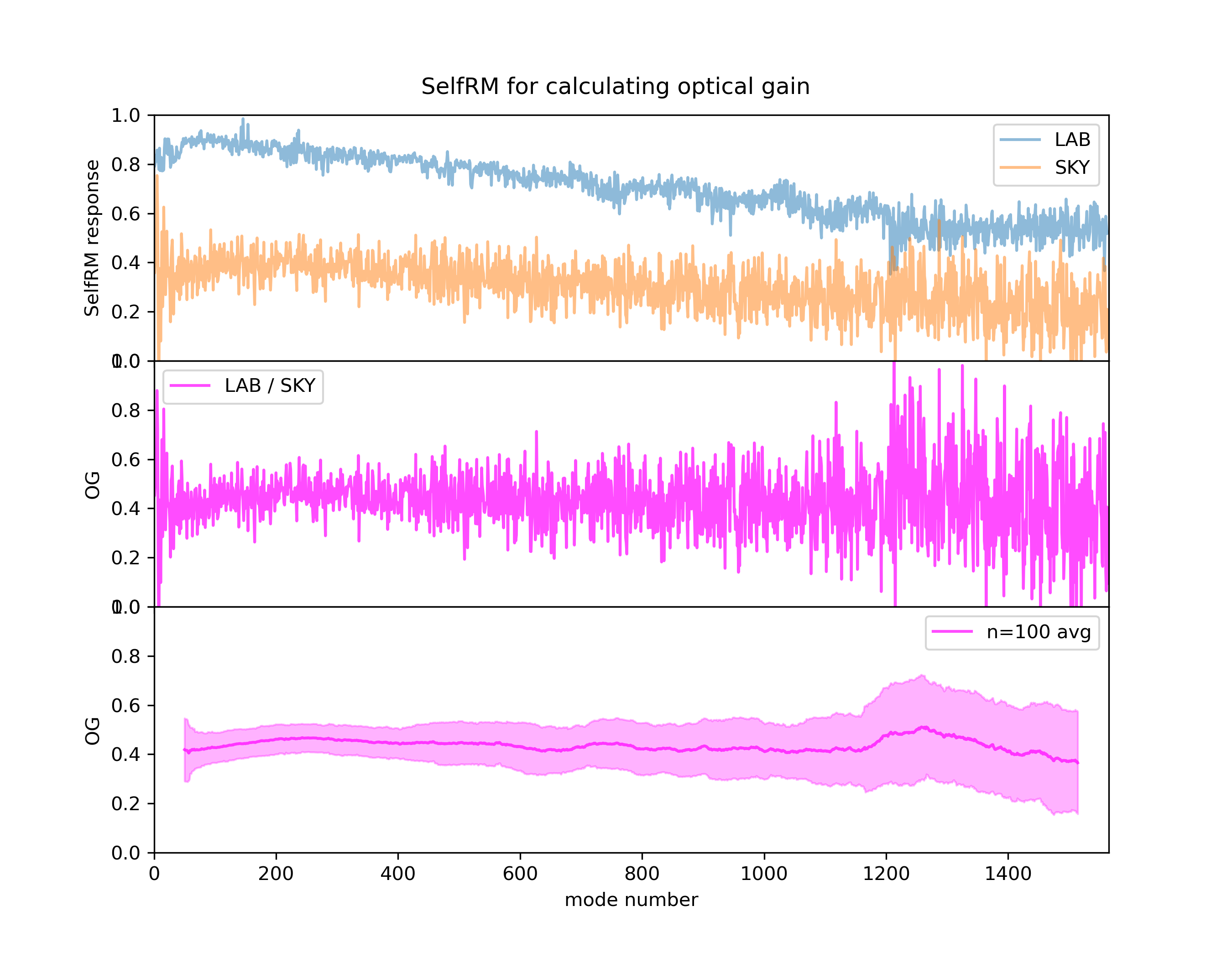}
    \caption{SelfRM for lab and on sky used to calculate modal optical gains. (Top) the selfRM modal response. (Middle) Division of the SKY by the LAB response for an optical gain curve. (Bottom) Optical gain curve smoothed by a 100 frame rolling average, showing a smooth response curve. }
    \label{fig.selfRM_OG}
\end{figure}

The selfRM when plotted by mode, for both an example sky and lab acquisition taken with MagAO-X, is shown in Figure \ref{fig.selfRM_OG}. The lab response does not maintain a flat 1.0 signal as selfRMs retain the scaling factors incorporated into the control matrix. These scaling factors are the same in lab and on sky and divide out. By dividing the sky response by the lab response an optical gain modal curve is recovered. When smoothed, this curve is approximately flat vs. mode number. 

\section{Optical Gain measured on sky}
\label{sect:OG}

OG curves using the selfRM method were collected through the MagAO-X 2022B and 2023A observing runs, for a total of 32 measurements. In the following section we detail key insights these measurements have allowed us to make on OG behavior on sky.

\subsection{Optical Gain vs. Mode number}
\begin{figure}[ht]
    \centering
    \includegraphics[width=0.95 \textwidth]{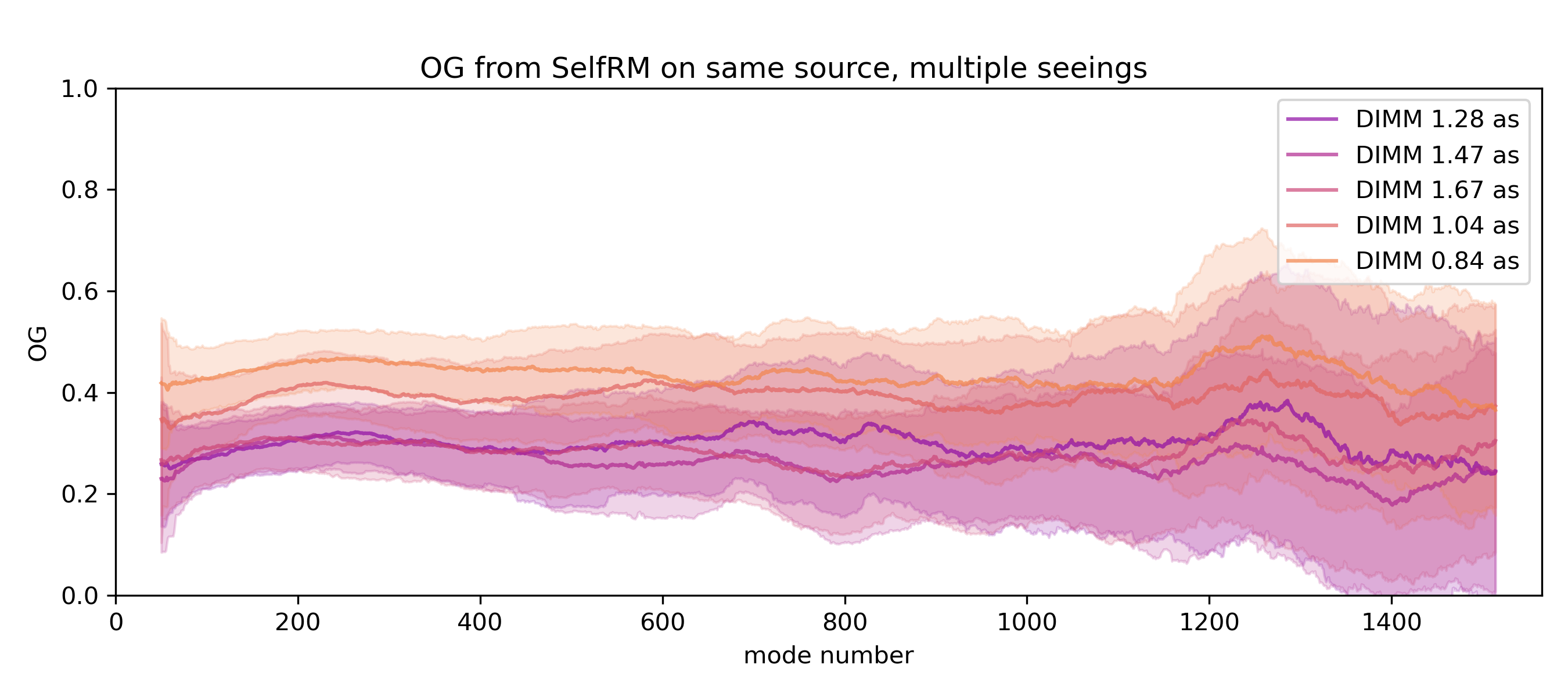}
    \caption{Multiple OG curves taken from the same target across varying seeing. When compared to closest DIMM seeing measurement scaled to PyWFS wavelength, an increase in seeing correlates to a reduction in OG.}
    \label{fig.selfRM_DIMM_some}
\end{figure}

On the same target with the same number of control modes but with seeing varying, Figure \ref{fig.selfRM_DIMM_some} shows that the selfRM retains its shape as the OG curve shifts to worse conditions. From this result, we feel confident that the optical gain response of the MagAO-X system is approximately flat to the degree necessary for effective OG control systems. 

\subsection{Optical Gain vs. DIMM Seeing}
\begin{figure}[ht]
    \centering
    \includegraphics[width=0.95 \textwidth]{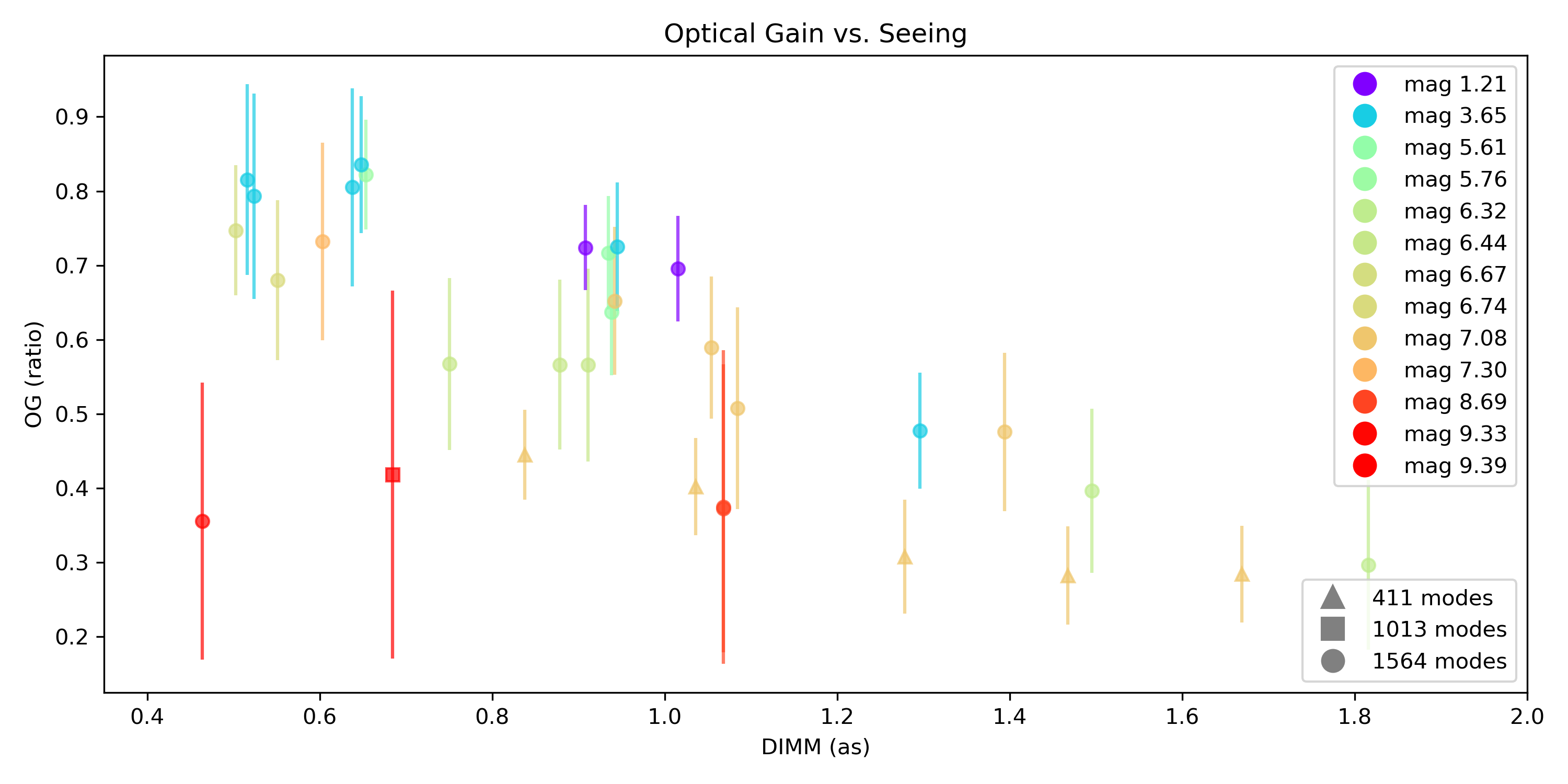}
    \caption{OG points for targets observed with MagAO-X, against the DIMM seeing measurement. The modes controlled at the time as well as the guide star magnitude are noted as these both influence system's residual WFE.}
    \label{fig.selfRM_DIMM}
\end{figure}

Figure \ref{fig.selfRM_DIMM} displays all measured OG points against DIMM seeing. The OG points displayed are calculated as the average of the modal OG curve, excluding the modes offloaded to the woofer (0-107) and the higher order modes with the greatest scatter (1207 to 1564). The error displayed is not uncertainty in OG but the measurement scatter across the remaining modes. The OG values shown in Figure \ref{fig.selfRM_DIMM} follow expected performance\cite{vanGorkom2021} on bright stars and good seeing, about 0.8. Guide star magnitude is recorded in the PyWFS wavelength of I band, calculated from Gaia Magnitudes\cite{Gaia_2016, Gaia_2023j}, using SIMBAD\cite{SIMBAD_2000} reported stellar types and the Mamajeck \cite{Mamajek2013} table for conversion. Performance decreases for fainter guide stars and worse seeing. 

\subsection{Optical Gain vs. Strehl Ratio}
\begin{figure}[ht]
    \centering
    \includegraphics[width=0.95 \textwidth]{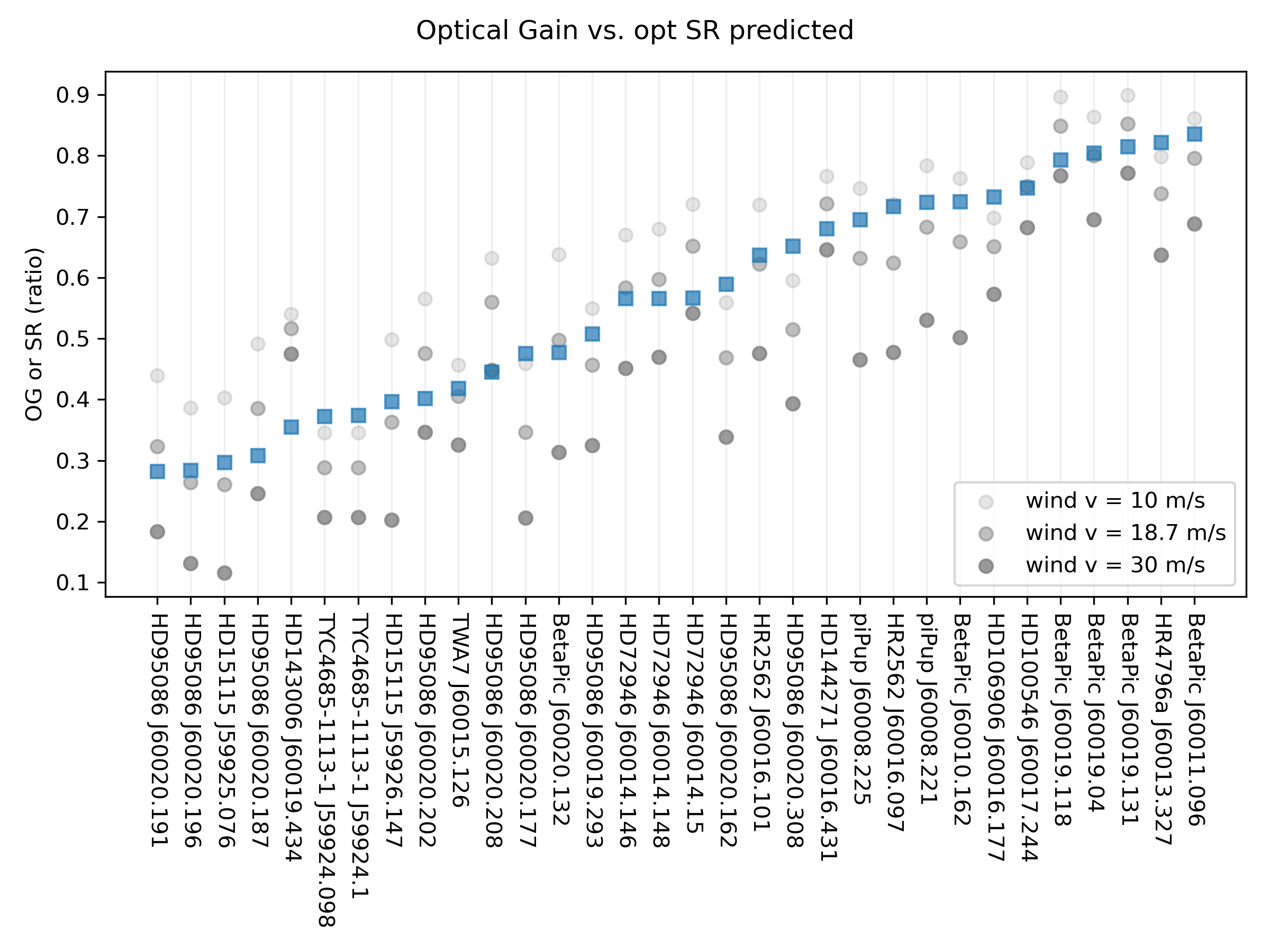}
    \caption{Strehl Ratio in gray, calculated using the telemetry of each observation. OG values sorted in ascending order, measured with selfRMs. OG values measured from selfRMs fall between SRs of the highest and lowest $v_{wind}$ speed for most observations.}
    \label{fig.SR_perOG}
\end{figure}

For a fuller understanding of our system's performance relative to OG we predict the Strehl Ratio (SR) at the tip of the PyWFS using telemetry from each observation. In Figure \ref{fig.SR_perOG} the SR calculated from the telemetry for each OG point is plotted against the OG value itself, and shows good agreement. In our forthcoming paper we describe how we predict SR and validate this result with simulations.

\section{Realtime Optical Gain tracking}
\label{sect:speckles}

With this preliminary work that confirms OG as flat as a function of mode number on the MagAO-X instrument, we have pursued correction algorithms that track optical gain at a single spatial frequency, following the work done at the LBT\cite{Esposito2020}. If a mode of known amplitude is applied to the DM throughout an observation, that mode's reconstruction amplitude, averaged over turbulence, can be monitored to track optical gain. 

\begin{figure}[ht]
    \centering
    \includegraphics[width=0.95 \textwidth]{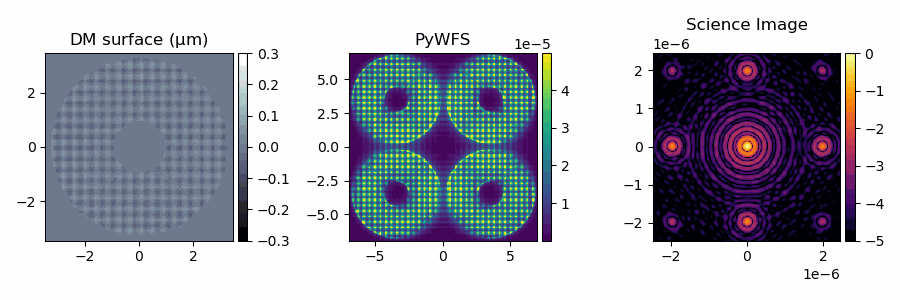}
    \caption{Frame of incoherent speckles in simulation. }
    \label{fig.sparkle}
\end{figure}

Incoherent Speckles\cite{Jovanovic_2015} are a high frequency probe already in routine use on the MagAO-X instrument. The Fourier mode applied in both x and y in the DM's pupil plane creates a PSF copy on the focal plane at separations of 10-22 $\lambda /D$ that allows for photometric calibration and centering during coronagraphic observations. A simulation of its effect on the DM, WFS, and focal plane is shown in Figure \ref{fig.sparkle}. We have pursued several methods of extracting the known amplitude pattern shape from WFS images, involving averaging over turbulence and comparing to lab reference images. Signals are present in the data extracted, but detailed optical gain tracking is stalled until real time control upgrades allow precise synchronization between the WFS camera frame and the DM command.

\section{Conclusion}
\label{sect:conclusion}

This work explored optical gain measurements on the MagAO-X instrument, in furthering the work of real time optical gain calibration and control. We have demonstrated:
\begin{itemize}
    \item Optical Gain is approximately flat with respect to mode number.
    \item Optical Gain follows the Strehl Ratio of the PSF on the tip of the pyramid.
\end{itemize}

Through a selfRMs, a CACAO process for extracting modal response amplitudes, we have measured optical gain values across 0.4 to 1.8as DIMM seeing across target magnitude 1.2 to 9.4 magnitudes. This sample has allowed us to confirm that optical gain is approximately flat across the MagAO-X modal control basis. 

Using a simple error budget and assumed wind speed, Strehl ratios were calculated for each OG measurement. Comparing estimated SR to measured OG show that OG follows SR on sky as predicted in simulations.

Current unfinished work has been undertaken to measure OG with the incoherent speckles already native to the MagAO-X control scheme. With the conclusions derived here we are confident that real time measurement and control of OG is possible with a high frequency DM probe.

\acknowledgments 
 
We are very grateful for the support from the NSF MRI Award MRI Award \#1625441 for MagAO-X. Eden McEwen would like to acknowledge her funding through the NSF GRFP. Thank you to the MagAO-X observers and collaborators who gave time for SelfRM data collection from their observing time. 
This research made use of HCIPy, an open-source object-oriented framework written in Python for performing end-to-end simulations of high-contrast imaging instruments (Por et al. 2018).
This work has made use of data from the European Space Agency (ESA) mission {\it Gaia} (\url{https://www.cosmos.esa.int/gaia}), processed by the {\it Gaia} Data Processing and Analysis Consortium (DPAC, \url{https://www.cosmos.esa.int/web/gaia/dpac/consortium}). Funding for the DPAC has been provided by national institutions, in particular the institutions participating in the {\it Gaia} Multilateral Agreement. 
This research has made use of the SIMBAD database, operated at CDS, Strasbourg, France.  

\bibliography{report} 
\bibliographystyle{spiebib} 

\end{document}